\documentclass{article}

\usepackage[english]{babel}

\usepackage[letterpaper,top=2cm,bottom=2cm,left=3cm,right=3cm,marginparwidth=1.75cm]{geometry}

\usepackage{amsmath}
\usepackage{graphicx}
\usepackage[colorlinks=true, allcolors=blue]{hyperref}
\usepackage{multirow}
\usepackage{natbib}

\usepackage{authblk}
\usepackage{graphicx}

\begin{document}

\title{News and Misinformation Consumption in Europe:\\ A Longitudinal Cross-Country Perspective}
\author[1]{Anees Baqir}
\author[1]{Alessandro Galeazzi}
\author[1,2,*]{Fabiana Zollo}

\affil[1]{Ca' Foscari University of Venice, Italy}
\affil[2]{The New Institute Centre for Environmental Humanities, Italy}
\affil[*]{Correspondence to: fabiana.zollo@unive.it}

\date{}

\maketitle

\begin{abstract}
The Internet and social media have transformed news availability and accessibility, reshaping information consumption and production. However, they can also facilitate the rapid spread of misinformation, posing significant societal challenges. To combat misinformation effectively, it is crucial to understand the online information environment and news consumption patterns.
Previous studies have shown that online debates often exhibit high levels of polarization intertwined with misinformation. Most existing research has primarily focused on single topics or individual countries, lacking cross-country comparisons. This study investigated information consumption in four European countries, focusing on the role of misinformation sources and analyzing three years of Twitter activity from news outlet accounts in France, Germany, Italy, and the UK. 
Furthermore, our work offers a perspective on how topics of European significance are interpreted across various countries.
The results indicate that reliable sources largely dominate the information landscape, although unreliable content is still present across all countries and topics. While most users engage with reliable sources, a small percentage consume questionable content. Interestingly, few users have a mixed information diet, but they bridge the gap between questionable and reliable news in the similarity network.
Cross-country comparisons revealed differences in audience overlap of news sources, offering valuable guidance for policymakers and scholars seeking to develop effective and tailored solutions to combat misinformation. Measuring the presence of misinformation and understanding its consumption dynamics is essential for tackling the challenges posed by the swift dissemination of unreliable information in online spaces.
\end{abstract}

\section{Introduction}
The advent of the Internet has revolutionized how we access information, granting users the capacity to engage directly with content and receive real-time feedback, reshaping the information landscape and presenting both opportunities and challenges. A primary concern is the potential rapid dissemination of misinformation and its far-reaching impact on various aspects of society, spanning from the realm of politics~\citep{stella2018bots,del2017mapping,bovet2019influence,flamino2023political,ferrara2017disinformation,grinberg2019fake}, to critical societal issues like climate change~\citep{falkenberg2022growing} and vaccines~\citep{schmidt2018polarization,santoro2023analyzing}. The presence of misinformation on social media has been acknowledged as a phenomenon with the potential to influence the outcomes of crucial societal processes, leading scholars to increasingly focus on addressing this issue. As a response, extensive discussions involving scholars and policymakers have been centered on strategies to mitigate the spread of misinformation, including recent legislative initiatives within the European Union aimed at compelling social media platforms to implement countermeasures~\citep{eulaw}.

In recent years, a plethora of research has been dedicated to understanding the dynamics and factors that may influence the spread of misinformation~\citep{ruths2019misinformation}. Some studies have compared the dissemination patterns of reliable and questionable content in various contexts, including science and conspiracy theories~\citep{del2016spreading, zannettou2018gab, lazer2018science}, the Covid-19 pandemic~\citep{ferrara2020misinformation, cinelli2020covid}, vaccines~\citep{broniatowski2023efficacy,santoro2023analyzing}, and elections~\citep{grinberg2019fake}, revealing differences in diffusion dynamics and prominence between reliable and unreliable news sources. Researchers have also investigated the role of the information environment in the spread of misinformation, underscoring how polarized debates can create fertile ground for its dissemination ~\citep{garimella2021political}. Echo chambers, where like-minded individuals reinforce their beliefs through repeated interactions, have been explored, indicating that misinformation primarily circulates within specific user groups~\citep{cinelli2021echo}. 
Furthermore, factors suspected of influencing news consumption may include social media recommendation algorithms, which can impact exposure to ideologically diverse news~\citep{flaxman2013ideological,bakshy2015exposure, nyhan2023like,gonzalez2023asymmetric}, and automated accounts,  which have been implicated in amplifying misinformation~\citep{stella2018bots,bessi2016social, zannettou2019disinformation}.

Although there is a substantial body of literature on misinformation, most studies have centered on individual countries or specific subjects. In this work, we took a distinct approach by conducting a comparative analysis of misinformation spanning various topics in diverse European countries. This approach enabled us to highlight the differences and similarities in interest, engagement, and consumption of information over time and across European countries. 

We investigated the consumption of Twitter content produced by news outlets in Europe, focusing on events from 2019 to 2022. Our goal was to offer a comparative assessment of the information landscape across multiple countries. To ensure a topic-independent analysis, we select one subject per year that has been debated in all four countries under consideration: France, Germany, Italy, and the United Kingdom. We analyzed the engagement generated within these countries and around these topics, while taking into account the reliability of the content sources. Furthermore, we constructed similarity networks based on the consumption patterns of news outlets' content, allowing us to compare the diverse structures that emerge across countries and topics.

Our findings revealed that reliable sources dominated the information landscape, though there was active participation from questionable user groups in the debate.  Notably, our networks indicated that users engage with both types of information sources. Furthermore, our cross-country comparison uncovered variations in the similarity structure of news sources among countries, ranging from a clear separation of questionable sources to a more mixed composition with no significant differences.

Overall, our results highlighted disparities as well as commonalities in news consumption among the chosen countries, especially concerning subjects of shared European interest, offering a valuable view of the topic perception across different European nations.
We also emphasized the role played by questionable sources, providing insights at both the country and topic levels that can be leveraged in the design of effective measures to counter misinformation.

\section{Materials and Methods}
\label{sec:methods}

\subsection*{Data collection and processing}
\label{subsec:dataset}

The data was collected using the official Twitter API for academic research~\footnote{\url{https://developer.twitter.com/en/docs/twitter-api}}, freely available for academics at the time of collection. Based on the list of accounts retrieved from the NewsGuard dataset (see Table \ref{tab:newsguard}), we downloaded the Twitter timelines of media sources based in Italy, Germany, France, and the UK over three years from 2019 to 2021.
NewsGuard is a tool that evaluates the reliability of news outlets based on nine journalistic criteria. Following such criteria, a team of professional and independent journalists assigns a ``trust score'' between 0 and 100 to each news outlet. Ratings are not provided for individuals, satirical content, or social media platforms like Twitter, Facebook, and YouTube. News sources are categorized into two groups based on their score: Reliable (trust score greater or equal to 60) and Questionable (trust score less than 60). The threshold is set by NewsGuard based on the evaluation criteria.

\begin{table}[h]
    \centering
    \begin{tabular}{c|c|c|c}
    \hline
    \hline
        \textbf{Country} & \textbf{Reliable sources} & \textbf{Questionable sources} & \textbf{Total} \\
        \hline
        France & 187 & 49 & 236 \\
        \hline
        Germany & 196 & 25 & 221 \\
        \hline
        Italy & 175 & 29 & 204 \\
        \hline
        UK & 191 & 22 & 213 \\
        \hline
        Total & 749 & 125 & 874 \\
        \hline
        \hline
    \end{tabular}
    \caption{ Breakdown of the NewsGuard news sources dataset by country and reliability}
    \label{tab:newsguard}
\end{table}

We collected only publicly available content from public Twitter accounts. The dataset included all the tweets published by the selected accounts in the period from 01 January 2019 to 11 November 2021, resulting in 25+ Million tweets. Table~\ref{tab:dataset} reports the breakdown of the data. The percentage of posts by each country contributing to the total amount is shown in parentheses. 

\begin{table}[h]
    \centering
    \begin{tabular}{c|c|c|c}
    \hline
    \hline
        \textbf{Country} & \textbf{Number of tweets} & \textbf{Reliable tweets} & \textbf{Questionable tweets} \\ \hline
        France & 7,083,659 (28.19\%) & 6,151,554 (26.57\%) & 932,105 (47.32\%) \\
        \hline
        Germany & 4,904,179 (19.52\%) & 4,689,186 (20.25\%) & 214,993 (10.91\%) \\
        \hline
        Italy & 4,936,407 (19.65\%) & 4,528,606 (19.56\%) & 407,801 (20.70\%) \\
        \hline
        UK & 8,201,352 (32.64\%) & 7,786,239 (33.62\%) & 415,113 (21.07\%) \\
        \hline
        Total & 25,125,597 & 23,155,585 & 1,970,012 \\
        \hline
        \hline
    \end{tabular}
    \caption{Volume of tweets by country and reliability}
    \label{tab:dataset}
\end{table}

To ensure that our analysis concentrated on topics debated at the European level for cross-country comparisons, we applied keyword filters to our original dataset. We divided our dataset into three one-year segments and filtered each segment according to a list of keywords related to the most discussed topic at the European level for that year. The statistics for the filtered data can be found in Table~\ref{tab:retweets}.
For the tweets in the filtered dataset, we collected all retweets. Details about the number of original tweets and retweets for each topic can be found in Table~\ref{tab:retweets}.

\begin{table}[h]
    \centering
    \resizebox{\textwidth}{!}{
    \begin{tabular}{p{2.34cm}|c|c|c|c|c|c|c}
    \hline
    \hline

     \multirow{6}{*}{\textbf{Brexit}} & \textbf{Keywords} & & \textbf{France} & \textbf{Germany} & \textbf{Italy} & \textbf{UK} & \textbf{Total} \\ \cline{1-8}
     & \multirow{4}{*}{brexit} &  Users & 33,288 & 22,512 & 8,676 & 231,911 & 296,387  \\ \cline{3-8}
      & & News sources & 129 & 127 & 89 & 167 & 512 \\ \cline{3-8}
      & & Tweets & 12,493 & 6,368 & 3,877 & 46,404 & 69,142 \\ \cline{3-8}
      & & Retweets & 97,909 & 53,352 & 24,856 & 1,385,023 & 1,561,140 \\ \hline

      \multirow{4}{*}{\textbf{Coronavirus}} & \multirow{4}{*}{\shortstack{ncov, \\ corona*, \\ covid*, \\ sars-cov-2}} & Users & 461,737 & 541,773 & 146,205 & 910,955 & 1,940,670 \\ \cline{3-8}
      & & News sources & 218 & 192 & 171 & 202 & 783 \\ \cline{3-8}
      & & Tweets & 204,728 & 99,527 & 117,899 & 137,913 & 560,067 \\ \cline{3-8}
      & & Retweets & 3,548,617 & 2,270,063 & 1,474,654 & 3,613,294 & 10,906,628 \\ \hline

      \multirow{4}{*}{\textbf{Covid Vaccine}} & \multirow{4}{*}{\shortstack{vacc*, astrazeneca,\\ vaccin*, moderna, \\ pfizer, sinopharm, \\ sputnik, biontech}} & Users & 396,131 & 165,122 & 156,273 & 303,365 & 1,020,891 \\ \cline{3-8}
      & & News sources & 214 & 180 & 171 & 192 & 757 \\ \cline{3-8}
      & & Tweets & 133,962 & 37,721 & 136,814 & 44,212 & 352,709 \\ \cline{3-8}
      & & Retweets & 2,630,179 & 779,521 & 1,943,585 & 1,099,068 & 6,452,353 \\
         
    \hline
    \hline
    \end{tabular}
    }
    \caption{Breakdown of the filtered dataset by Country and Topic.}
    \label{tab:retweets}
\end{table}

\subsection*{Similarity networks}
\label{subsec:network}
We assessed the audiences' similarity among news outlets exploiting the retweets of the content they produced.
For each country and topic, we built an undirected weighted graph $G$, in which nodes represent news outlets and edges the audience similarity among them. We started by creating a matrix $R_{c,t}$ for each country and each topic, with retweeters as rows and news outlets as a column, whereas $c \in$ \textit{\{France, Germany, Italy, UK\}} and $t \in$ \textit{\{Brexit, Coronavirus, Covid Vaccine\}}. The entry $r_{i,j}$ of $R_{c,t}$ is the number of times user $i$ retweeted a tweet posted by news source $j$ based in the country $c$ on topic $t$. We then computed the cosine similarity for each pair of columns to measure the audiences' similarity for each pair of news sources. Thus, the weight $w_{a,b}$ of the edge between node $a$ and $b$ in the graph $G$ is equal to:

\[w_{a,b} = \frac{r_a \cdot r_b}{\Vert r_a \Vert \Vert r_b \Vert}\]

where $r_a$ and $r_b$ are the two column vectors of news sources $i$ and $j$, respectively. It should be noted that $w_{a,b}$ $\in$ [0, 1] since all the entries of the matrix are non-negative.

Finally, we excluded all the 0-degree nodes and deleted all the edges with a weight below the median of all edge weights. This approach enabled us to capture the strongest similarities among news outlets' audiences related to the selected topics within the European context.

\subsection*{Topic modeling}
\label{subsec:bertopic}

We utilized BERTopic, a topic modeling tool that extracts latent topics from a collection of documents, to identify the heated topics prevalent in all the countries under examination. BERTopic is a top2vec model generalized for pretrained sentence transformers~\citep{grootendorst2022bertopic} that has recently demonstrated promising results in various tasks. BERTopic generates coherent clusters of documents through three steps: 1) extracting document embeddings; 2) clustering embeddings; 3) creating topic representations using class-based TF-IDF \citep{sammut2011encyclopedia} (c-TF-IDF). In the first step, any pre-trained transformer-based language models can be utilized, allowing the use of state-of-the-art embedding techniques. The second step employs uniform manifold approximation and projection (UMAP) to reduce the dimension of embeddings~\citep{mcinnes2018umap}, and hierarchical density-based spatial clustering of applications with Noise (HDBSCAN) to generate semantically similar clusters of documents~\citep{mcinnes2017hdbscan}. One of the topics is set to be ‘others’, and includes the documents that are not included in different topics. 
\section{Results and Discussion}
\label{sec:results}
In this section, we present the results of our analysis, organized as follows. First, we provide an overview of the information landscape in selected European countries over the three years. This step is crucial for identifying key topics that are widely shared among countries and distinguishing between questionable and reliable sources, enabling a coherent comparison.
Next, we examine both commonalities and differences among countries in their online discussions of these topics, focusing on user engagement and consumption patterns.

\subsection{The evolution of Public Discourse across Countries}
\begin{figure}[h]
    \centering
    \includegraphics[width=\textwidth]{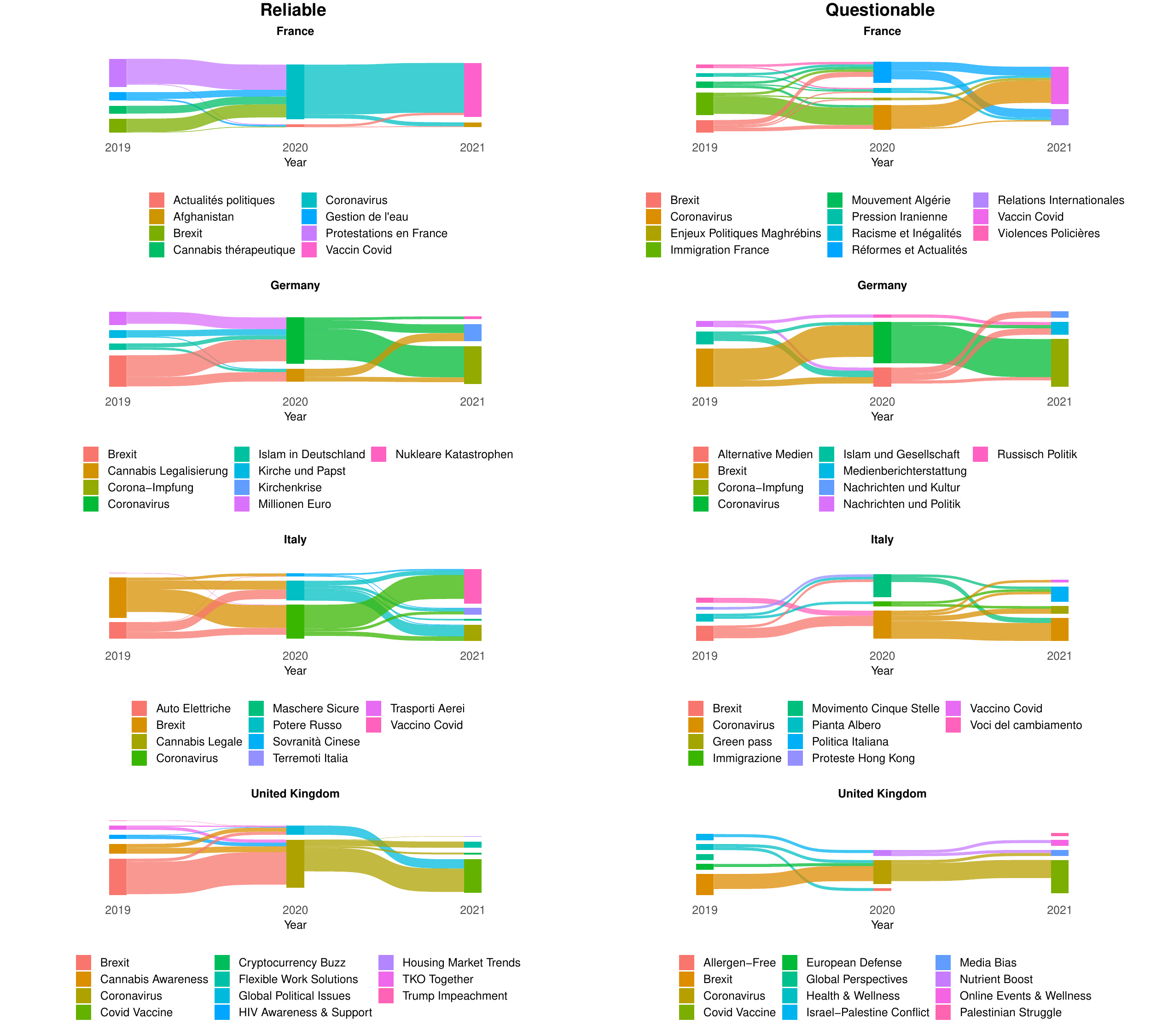}
    \caption{Topic modeling results on questionable and reliable news sources content across countries. The size of each topic is given by the proportion of unique news sources contributing to it. The flows represent the interest shift of news outlets in different topics over time.
    }
    \label{fig:topics}
\end{figure}

To compare the landscapes of public discourse in the selected countries, our initial step involves identifying common topics extensively discussed in all four countries and by both questionable and reliable sources. To this aim, we employ BERTopic\citep{grootendorst2022bertopic} to perform topic modeling on the content produced by news outlets' accounts over a three-year period (see Section \ref{sec:methods} for further details). To identify suitable topics for our analysis, we divide the dataset by year and by country and run BERTopic algorithm on each subset. The results reported in Figure~\ref{fig:topics} show the most debated topics for each year by country and source category. The size of each topic represents the number of news sources contributing to it, while its position reflects its relevance to the overarching topics. The flow diagrams show the topic's prevalence in news outlets over time.

Figure~\ref{fig:topics} highlights how the attention of news outlets to different topics varied across countries and types of news sources. Notably, in addition to certain topics of common interest, news outlets tended to prioritize subjects of national relevance, such as protests, the influence of foreign countries, religion, electric cars, and drug legalization. 
We also observe disparities in the topics covered by questionable and reliable sources within the same country. For instance, the fraction of news outlets reporting on the coronavirus vaccine in Italy was higher for reliable sources than for questionable ones. Furthermore, certain topics were exclusive to one type of source, like "Flights" (Italy, reliable), "Water management" (France, reliable), or "Palestinian struggle" (UK, questionable). These findings indicate that the level of interest was influenced both by the country and the type of source considered, with questionable sources displaying a broader range of interests and reliable ones focusing more on topics common to all countries.

Crucially, our analysis highlights the presence of common topics between both questionable and reliable debates of all countries. Specifically, three topics appeared consistently in debates across all countries: ``Brexit''(2019), ``Coronavirus''(2020), and ``Covid Vaccine''(2021). 
Therefore, in the subsequent analysis, we exclusively focus on these topics for a cross-country examination of the discourse. The rationale behind this choice is to spotlight the differences and similarities in how these topics were reported and consumed by news outlets and users from various countries, thereby minimizing the impact of topic-specific variations on our analysis. Additionally, these topics have been extensively discussed at the European level, making our analysis valuable for understanding how subjects of European significance are perceived across different countries.

\begin{figure}[h]
    \centering
    \includegraphics[width=\textwidth]{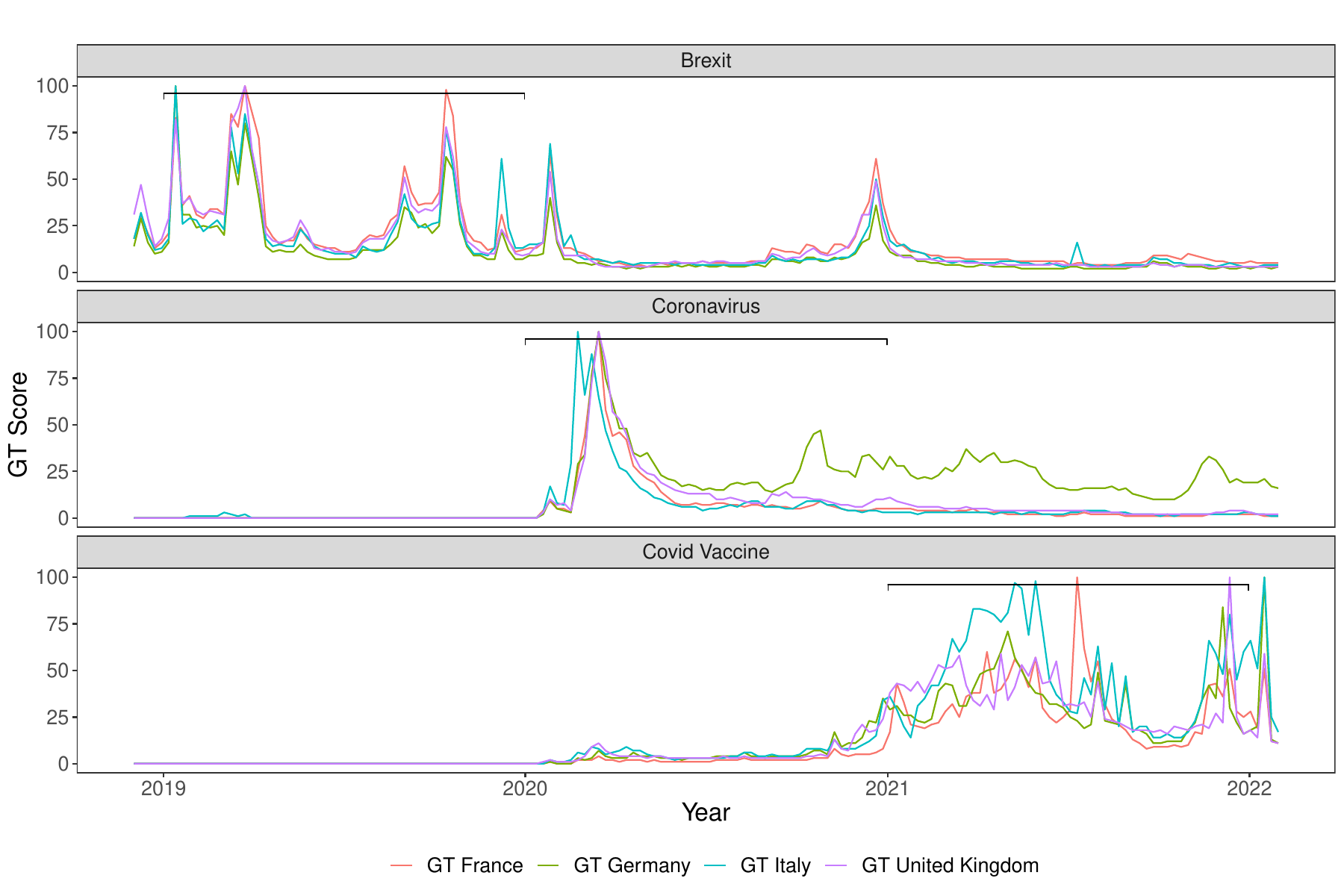}
    \caption{Google Trends analysis of search interest in Brexit, Coronavirus, and Covid Vaccine in France, Germany, Italy, and UK from 2019 to 2021. 
    The plots display how search interest for each topic evolved over time, with each row representing one topic. Interest trends reveal that Brexit was most popular in 2019, followed by a sharp decline in 2020 and 2021 with some exceptions at the end of 2020. Coronavirus peaked in early 2020 and declined thereafter, while Covid Vaccine gained momentum in early 2021, reached the maximum in mid-2021, and saw another surge at the end of 2021. Brackets represent the time span taken into account in the analysis for each topic.}
    \label{fig:gt}
\end{figure}

To underscore the relevance of the three chosen topics in online public debates and validate the accuracy of the time frames assigned to each topic, we conduct a Google Trends analysis of search interest in Brexit, Coronavirus, and Covid Vaccine in France, Germany, Italy, and the UK from 2019 to 2021, as shown in Figure \ref{fig:gt}.

The analysis of Google Trends confirms that the selected topics attracted the highest attention during the specified time frames in the broader online context. Thus, going forward, our analysis focuses on these three topics (Brexit, Coronavirus, and Covid Vaccine) to examine the differences and similarities in news production and consumption within the European landscape. To conduct our analysis exclusively on these topics, we filter the timelines of news outlets to select only tweets relevant to the chosen topic within the respective time range (see Section~\ref{sec:methods} for details).

\subsection{User engagement and community structures}
\begin{figure}[h]
    \centering
    \includegraphics[width=\textwidth]{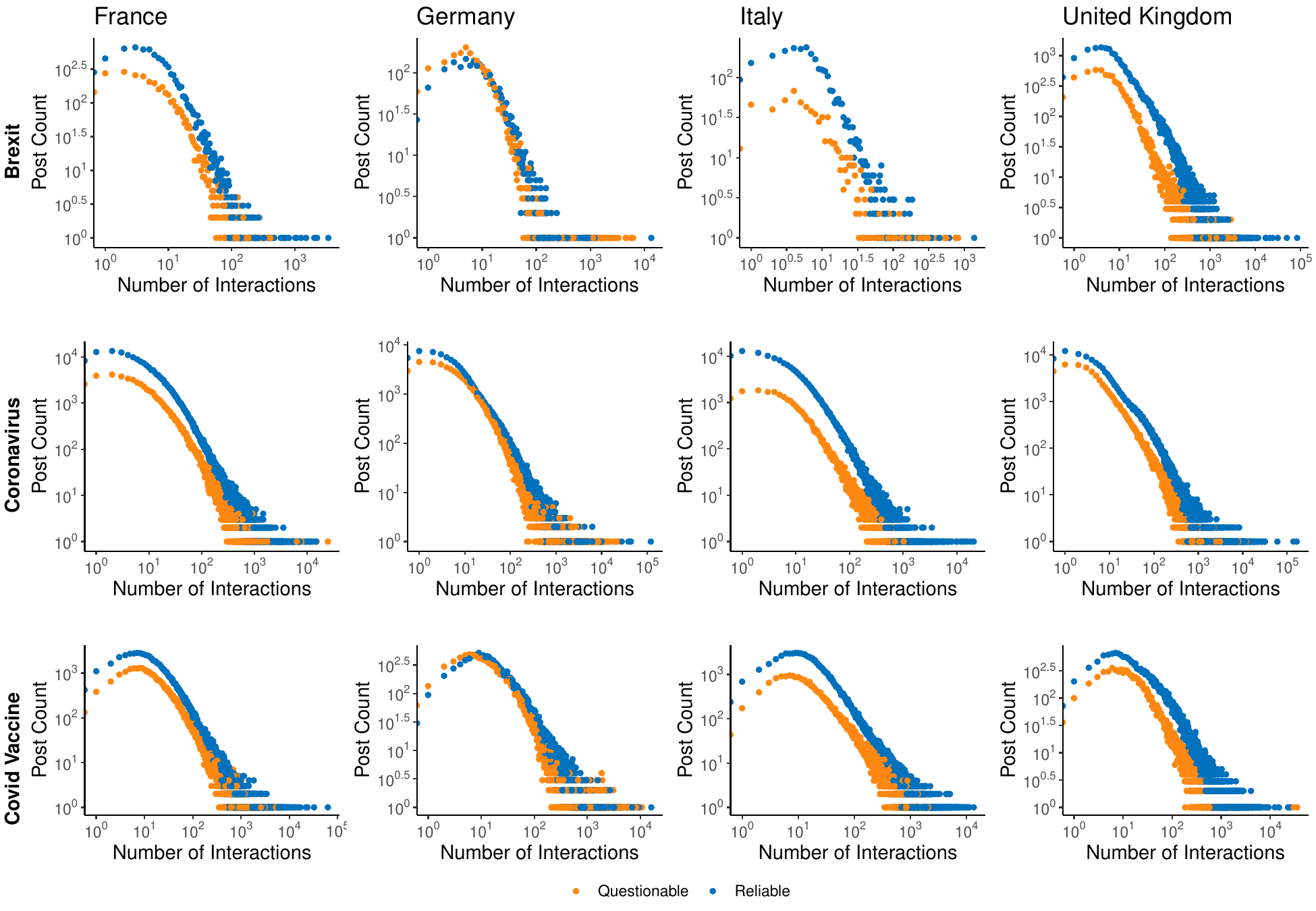}
    \caption{Distribution of tweet interactions by country for reliable (blue) and questionable (orange) news sources around Brexit (top row), Coronavirus (middle row), and Covid Vaccine (bottom row). Tweet interactions are computed as the sum of likes, retweets, quotes, and comments received by each tweet.} 
    \label{fig:engagement}
\end{figure}

We continue our study by comparing the engagement with content related to the identified topics on social media platforms.
Figure \ref{fig:engagement} shows the distribution of tweet interactions by country, computed as the sum of likes, retweets, quotes, and replies, for reliable news sources (blue) and questionable news sources (orange), as classified by NewsGuard (see Section~\ref{sec:methods}), for each of the three topics.
Despite minor geographical variations, the distributions of user interactions display a similar long-tailed distribution for all three topics, where a small number of tweets receive a large number of interactions while the majority receive very few.
Reliable news sources typically obtained more interactions than questionable sources, as shown by their wider distribution along the x-axis. However, a few exceptions are observed, such as the case of the UK in COVID-19 vaccine discussions and France in Coronavirus debates. Furthermore, in the Brexit discourse, questionable sources have a notable presence in the tail of the distribution in Germany and Italy, although they are less prominent in other discussions.
Overall, the presence of questionable sources and the engagement they generated can vary, contingent on both the country and the specific topic under consideration. 

We then turn our attention to news consumption patterns to highlight the differences and similarities in the news outlets' audiences. 
Analyzing Twitter data on Brexit, Coronavirus, and Covid Vaccine, we explore whether news outlets of the same type are consumed by similar audiences. We define a metric based on cosine similarity(see Section~\ref{sec:methods}) on retweeters to quantify the similarity between news outlets in terms of audiences. News outlets sharing a high percentage of retweeters have a higher value of the similarity metric (close to 1), while outlets with only a few shared retweeters get a low similarity (close to 0).

\begin{figure}[ht!]
    \centering
    \includegraphics[width=\textwidth]{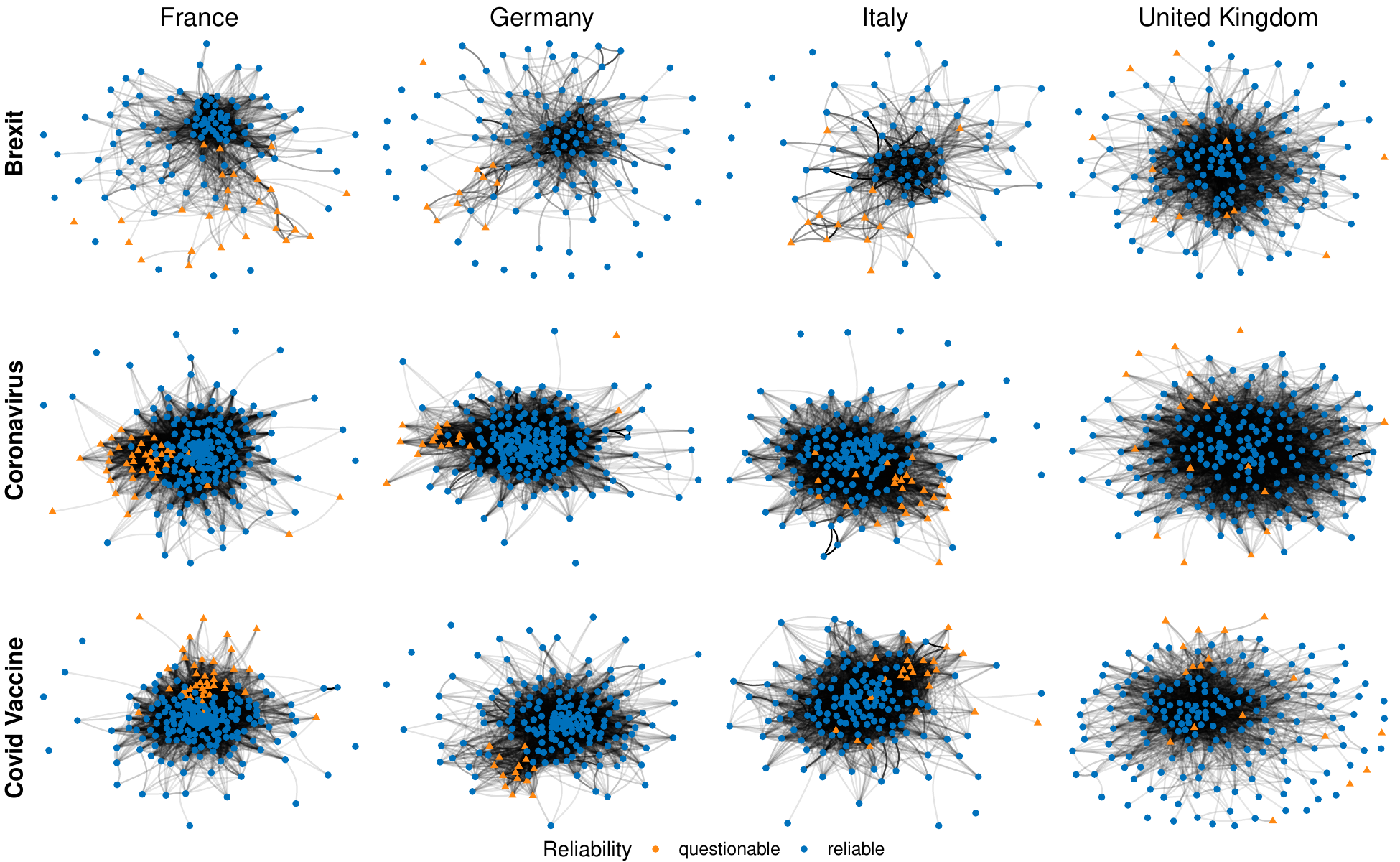}
    \caption{Similarity network among news outlets, where each news source is represented as a node, and edges represent audiences' similarity among news outlets. The color and shape of the nodes indicate the classification of the news source, and the thickness of the edges represents the level of similarity of retweeters between two news sources. We discarded edges with weights lower than the overall median of the edges. 
    Each network represents the news outlets' similarity on one topic for one country.}
    \label{fig:network}
\end{figure}

We then build an undirected network in which news outlets are represented as nodes and weighted edges indicate the level of similarity among them. We create one network for each country and topic considered to enable a fair comparison. The resulting networks are visualized in Figure \ref{fig:network}. To highlight only the stronger connections, we discard edges with weights lower than the overall median of the edges of each network (see Figures 1 and 2 of SI for the results with the complete networks).

We may observe variations in the network structure depending on the country and topic under consideration. Indeed, France, Germany, and Italy tend to display a clearly identifiable cluster of questionable sources (orange triangles), indicating the presence of communities primarily consuming questionable content. In the UK, this distinction is less pronounced.
Looking at topic-specific differences, we find that for all countries except the UK, the networks tend to be sparser, with a lower edge density, in the case of Brexit. For Coronavirus and Covid Vaccine discussions, the networks are more connected and exhibit higher edge density (see Table 2 of SI). This is reflected in the separation between questionable and reliable news sources: in the Brexit debate, the separation between the two types of news appears clearer, while in the other debates, they share a higher number of connections, as shown in Table 3 of SI. 
To quantify this behavior further, we apply the adjusted nominal assortativity to our networks~\citep{karimi2022inadequacy}, showing that higher levels of assortativity are achieved in the context of the Brexit debate. However, the UK exhibits different behavior, possibly due to its direct involvement in the debate.

\begin{figure}[h]
    \centering
    \includegraphics[width=\textwidth]{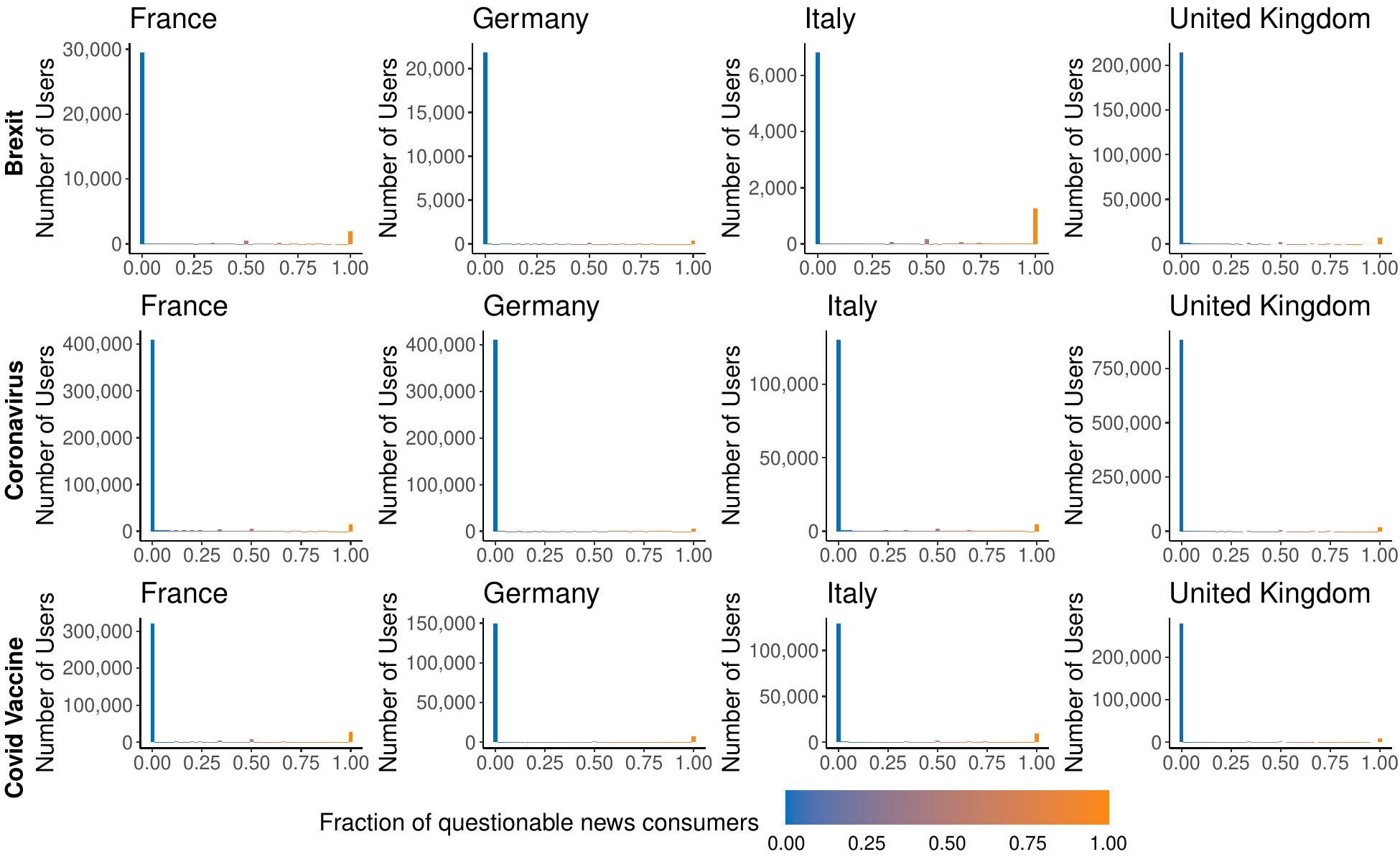}
    \caption{Analysis of user content consumption where each histogram represents the user count versus the fraction of news from potentially questionable sources, ranging from entirely reliable (0) to entirely questionable (1). A dominant presence near lower fractions suggests a prevalent reliance on reliable sources. In contrast, significant increases near the higher end highlight segments influenced by questionable content. }
    \label{fig:news_dist}
\end{figure}

\begin{figure}[h]
    \centering
    \includegraphics[width=\textwidth]{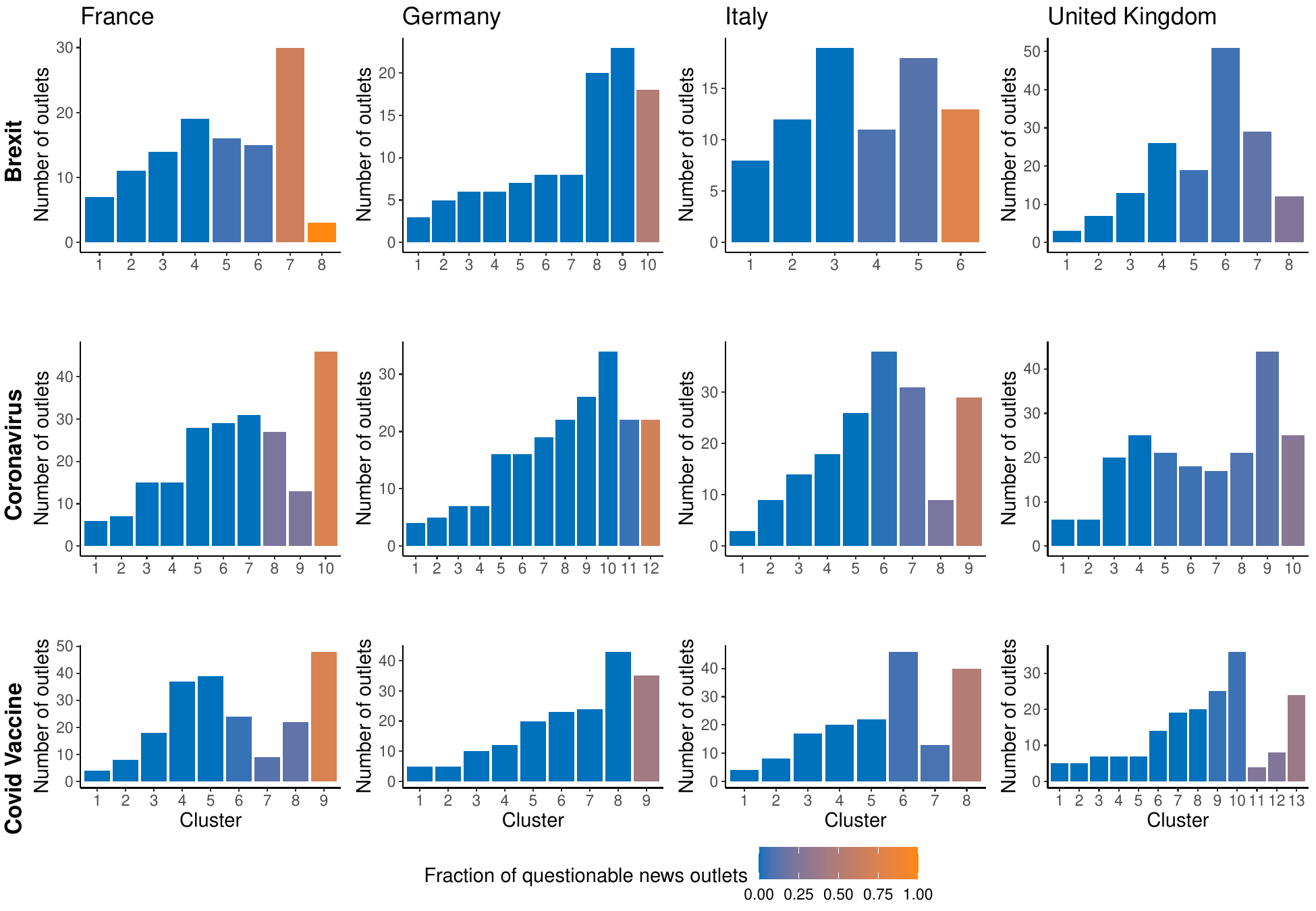}
    \caption{Community detection analysis of news outlets' similarity networks. Clusters were found using the Louvain clustering  algorithm and sorted based on the percentage of questionable news outlets. The percentage of questionable sources in each cluster is color coded. Network edges with weights lower than the median value were discarded here, result with the complete network is reported in SI.}
    \label{fig:communities}
\end{figure}

Our analysis also reveals that there is no absolute separation between questionable and reliable news outlets. This suggests that some users primarily or exclusively consume reliable or questionable content, while others have a mixed news diet, consuming both types in varying proportions. 
To delve deeper into this question, we analyze the fraction of questionable news consumed by each user and present the distribution in Figure~\ref{fig:news_dist}.
The results indicate that the majority of users in each debate primarily rely on reliable news sources (see also Table 4 of SI). However, in every debate, there is a small but noticeable fraction of users who exclusively endorse questionable news, although with varying degrees of prominence.
Notably, the Figure depicts a distinctive bimodal distribution, with very few users falling outside the extreme ends of the spectrum. These users play a crucial role in bridging the gap between questionable and reliable news within the similarity networks.
Furthermore, reliable news sources tend to occupy the core of the network, while questionable sources are generally situated in more peripheral positions. Indeed, among the top 25 sources identified by the PageRank algorithm in each network~\citep{bakshy2011identifying}, a substantial majority (at least 95.3\%) are found to be reliable news sources (see SI for further details).
We conclude our analysis by examining the community structure of the similarity networks. We perform community detection using the Louvain clustering algorithm \citep{blondel2008fast} and report the results in Figure~\ref{fig:communities}. Clusters are color-coded based on the proportion of questionable news outlets, with darker shades indicating a higher percentage of questionable sources.

Across all countries and topics, the majority of clusters consisted mainly of reliable news outlets, and within these clusters, we also find the most significant nodes according to the PageRank classification.  
However, our analysis also reveals the presence of small clusters with a high proportion of questionable news outlets. The number and size of these clusters vary depending on the country and topic. For instance, in Germany and Italy, there is one such cluster for each topic, while in the Brexit debate in France, there are two clusters. In the UK, the separation is less clear, with no clusters showing a high percentage of questionable news outlets.
We also notice that reliable clusters tend to be smaller in size but more numerous, while questionable clusters tend to be larger and often unique in each network. This suggested that users who consume questionable content tend to endorse most of the questionable sources of the network, while reliable news consumers focus on fewer news outlets.

Overall, our analysis provides a longitudinal view of the online news consumption landscape in the selected countries, highlighting the predominance of reliable news sources while also revealing the presence of clusters with a higher proportion of questionable news sources in many countries and topics. The existence of such clusters suggests the presence of a group of users consuming content from various questionable sources while avoiding reliable ones. This behavior is consistent with the potential presence of echo chambers, a phenomenon widely observed in online debates~\citep{cinelli2021echo,falkenberg2022growing,cota2019quantifying}.

\section{Conclusions}
In this study, we have delved into the evolving dynamics of news production and consumption within the European context. We examined the consumption of Twitter content produced by news outlets in France, Germany, Italy, and the United Kingdom, providing a cross-country and cross-topic comparison of the online public discourse. We identified topics debated across all four countries and highlighted differences and similarities in consumption patterns. Additionally, we constructed networks based on the similarities among news outlets' audiences, revealing the presence of groups of users engaging with sources of different reliability.

Our findings indicated that reliable sources dominate the information landscape, but users consuming content mainly or exclusively from questionable news outlets were often present. However, the size and importance of such groups vary based on the topic and the country under consideration. Furthermore, our cross-country comparison has revealed variations in the structure of news sources' similarity networks. While some countries exhibited a clearer separation between clusters of questionable sources and reliable sources, others showed a more heterogeneous situation with less detectable differences in cluster composition. However, the connectedness of the networks and users' behavior analysis indicated the presence of a small fraction of users with a mixed news diet in all countries.

Our results emphasized the differences and similarities in news consumption patterns across countries in relation to globally significant subjects. Understanding the dynamic of news consumption and its dependence on factors such as the topic or country can provide valuable insights into the development of effective countermeasures to mitigate the spread of misinformation and disinformation. Monitoring the information landscape at both national and European levels is indeed crucial to understanding the state of public discourse on contentious topics and developing tailored cohesive strategies to improve the health of information ecosystems.
 
\bibliographystyle{apa}
\bibliography{references}

\begin{thebibliography}{}

\bibitem[\protect\astroncite{eul}{}]{eulaw}
European commission, the digital services act package.
\newblock accessed on 23-10-2023.

\bibitem[\protect\astroncite{Bakshy et~al.}{2011}]{bakshy2011identifying}
Bakshy, E., Hofman, J.~M., Mason, W.~A., and Watts, D.~J. (2011).
\newblock Identifying influencers on twitter.
\newblock In {\em Fourth ACM International Conference on Web Seach and Data
  Mining (WSDM)}, volume~2.

\bibitem[\protect\astroncite{Bakshy et~al.}{2015}]{bakshy2015exposure}
Bakshy, E., Messing, S., and Adamic, L.~A. (2015).
\newblock Exposure to ideologically diverse news and opinion on facebook.
\newblock {\em Science}, 348(6239):1130--1132.

\bibitem[\protect\astroncite{Bessi and Ferrara}{2016}]{bessi2016social}
Bessi, A. and Ferrara, E. (2016).
\newblock Social bots distort the 2016 us presidential election online
  discussion.
\newblock {\em First monday}, 21(11-7).

\bibitem[\protect\astroncite{Blondel et~al.}{2008}]{blondel2008fast}
Blondel, V.~D., Guillaume, J.-L., Lambiotte, R., and Lefebvre, E. (2008).
\newblock Fast unfolding of communities in large networks.
\newblock {\em Journal of statistical mechanics: theory and experiment},
  2008(10):P10008.

\bibitem[\protect\astroncite{Bovet and Makse}{2019}]{bovet2019influence}
Bovet, A. and Makse, H.~A. (2019).
\newblock Influence of fake news in twitter during the 2016 us presidential
  election.
\newblock {\em Nature communications}, 10(1):7.

\bibitem[\protect\astroncite{Broniatowski
  et~al.}{2023}]{broniatowski2023efficacy}
Broniatowski, D.~A., Simons, J.~R., Gu, J., Jamison, A.~M., and Abroms, L.~C.
  (2023).
\newblock The efficacy of facebook’s vaccine misinformation policies and
  architecture during the covid-19 pandemic.
\newblock {\em Science Advances}, 9(37):eadh2132.

\bibitem[\protect\astroncite{Cinelli et~al.}{2021}]{cinelli2021echo}
Cinelli, M., De~Francisci~Morales, G., Galeazzi, A., Quattrociocchi, W., and
  Starnini, M. (2021).
\newblock The echo chamber effect on social media.
\newblock {\em Proceedings of the National Academy of Sciences},
  118(9):e2023301118.

\bibitem[\protect\astroncite{Cinelli et~al.}{2020}]{cinelli2020covid}
Cinelli, M., Quattrociocchi, W., Galeazzi, A., Valensise, C.~M., Brugnoli, E.,
  Schmidt, A.~L., Zola, P., Zollo, F., and Scala, A. (2020).
\newblock The covid-19 social media infodemic.
\newblock {\em Scientific reports}, 10(1):1--10.

\bibitem[\protect\astroncite{Cota et~al.}{2019}]{cota2019quantifying}
Cota, W., Ferreira, S.~C., Pastor-Satorras, R., and Starnini, M. (2019).
\newblock Quantifying echo chamber effects in information spreading over
  political communication networks.
\newblock {\em EPJ Data Science}, 8(1):35.

\bibitem[\protect\astroncite{Del~Vicario et~al.}{2016}]{del2016spreading}
Del~Vicario, M., Bessi, A., Zollo, F., Petroni, F., Scala, A., Caldarelli, G.,
  Stanley, H.~E., and Quattrociocchi, W. (2016).
\newblock The spreading of misinformation online.
\newblock {\em Proceedings of the national academy of Sciences},
  113(3):554--559.

\bibitem[\protect\astroncite{Del~Vicario et~al.}{2017}]{del2017mapping}
Del~Vicario, M., Zollo, F., Caldarelli, G., Scala, A., and Quattrociocchi, W.
  (2017).
\newblock Mapping social dynamics on facebook: The brexit debate.
\newblock {\em Social Networks}, 50:6--16.

\bibitem[\protect\astroncite{Falkenberg et~al.}{2022}]{falkenberg2022growing}
Falkenberg, M., Galeazzi, A., Torricelli, M., Di~Marco, N., Larosa, F., Sas,
  M., Mekacher, A., Pearce, W., Zollo, F., Quattrociocchi, W., et~al. (2022).
\newblock Growing polarization around climate change on social media.
\newblock {\em Nature Climate Change}, pages 1--8.

\bibitem[\protect\astroncite{Ferrara}{2017}]{ferrara2017disinformation}
Ferrara, E. (2017).
\newblock Disinformation and social bot operations in the run up to the 2017
  french presidential election.
\newblock {\em arXiv preprint arXiv:1707.00086}.

\bibitem[\protect\astroncite{Ferrara et~al.}{2020}]{ferrara2020misinformation}
Ferrara, E., Cresci, S., and Luceri, L. (2020).
\newblock Misinformation, manipulation, and abuse on social media in the era of
  covid-19.
\newblock {\em Journal of Computational Social Science}, 3:271--277.

\bibitem[\protect\astroncite{Flamino et~al.}{2023}]{flamino2023political}
Flamino, J., Galeazzi, A., Feldman, S., Macy, M.~W., Cross, B., Zhou, Z.,
  Serafino, M., Bovet, A., Makse, H.~A., and Szymanski, B.~K. (2023).
\newblock Political polarization of news media and influencers on twitter in
  the 2016 and 2020 us presidential elections.
\newblock {\em Nature Human Behaviour}, pages 1--13.

\bibitem[\protect\astroncite{Flaxman et~al.}{2013}]{flaxman2013ideological}
Flaxman, S., Goel, S., and Rao, J.~M. (2013).
\newblock Ideological segregation and the effects of social media on news
  consumption.
\newblock {\em Available at SSRN}, 2363701.

\bibitem[\protect\astroncite{Garimella et~al.}{2021}]{garimella2021political}
Garimella, K., Smith, T., Weiss, R., and West, R. (2021).
\newblock Political polarization in online news consumption.
\newblock In {\em Proceedings of the International AAAI Conference on Web and
  Social Media}, volume~15, pages 152--162.

\bibitem[\protect\astroncite{Gonz{\'a}lez-Bail{\'o}n
  et~al.}{2023}]{gonzalez2023asymmetric}
Gonz{\'a}lez-Bail{\'o}n, S., Lazer, D., Barber{\'a}, P., Zhang, M., Allcott,
  H., Brown, T., Crespo-Tenorio, A., Freelon, D., Gentzkow, M., Guess, A.~M.,
  et~al. (2023).
\newblock Asymmetric ideological segregation in exposure to political news on
  facebook.
\newblock {\em Science}, 381(6656):392--398.

\bibitem[\protect\astroncite{Grinberg et~al.}{2019}]{grinberg2019fake}
Grinberg, N., Joseph, K., Friedland, L., Swire-Thompson, B., and Lazer, D.
  (2019).
\newblock Fake news on twitter during the 2016 us presidential election.
\newblock {\em Science}, 363(6425):374--378.

\bibitem[\protect\astroncite{Grootendorst}{2022}]{grootendorst2022bertopic}
Grootendorst, M. (2022).
\newblock Bertopic: Neural topic modeling with a class-based tf-idf procedure.
\newblock {\em arXiv preprint arXiv:2203.05794}.

\bibitem[\protect\astroncite{Karimi and Oliveira}{2022}]{karimi2022inadequacy}
Karimi, F. and Oliveira, M. (2022).
\newblock On the inadequacy of nominal assortativity for assessing homophily in
  networks.
\newblock {\em arXiv preprint arXiv:2211.10245}.

\bibitem[\protect\astroncite{Lazer et~al.}{2018}]{lazer2018science}
Lazer, D.~M., Baum, M.~A., Benkler, Y., Berinsky, A.~J., Greenhill, K.~M.,
  Menczer, F., Metzger, M.~J., Nyhan, B., Pennycook, G., Rothschild, D., et~al.
  (2018).
\newblock The science of fake news.
\newblock {\em Science}, 359(6380):1094--1096.

\bibitem[\protect\astroncite{McInnes et~al.}{2017}]{mcinnes2017hdbscan}
McInnes, L., Healy, J., and Astels, S. (2017).
\newblock hdbscan: Hierarchical density based clustering.
\newblock {\em J. Open Source Softw.}, 2(11):205.

\bibitem[\protect\astroncite{McInnes et~al.}{2018}]{mcinnes2018umap}
McInnes, L., Healy, J., and Melville, J. (2018).
\newblock Umap: Uniform manifold approximation and projection for dimension
  reduction.
\newblock {\em arXiv preprint arXiv:1802.03426}.

\bibitem[\protect\astroncite{Nyhan et~al.}{2023}]{nyhan2023like}
Nyhan, B., Settle, J., Thorson, E., Wojcieszak, M., Barber{\'a}, P., Chen,
  A.~Y., Allcott, H., Brown, T., Crespo-Tenorio, A., Dimmery, D., et~al.
  (2023).
\newblock Like-minded sources on facebook are prevalent but not polarizing.
\newblock {\em Nature}, 620(7972):137--144.

\bibitem[\protect\astroncite{Ruths}{2019}]{ruths2019misinformation}
Ruths, D. (2019).
\newblock The misinformation machine.
\newblock {\em Science}, 363(6425):348--348.

\bibitem[\protect\astroncite{Sammut and Webb}{2011}]{sammut2011encyclopedia}
Sammut, C. and Webb, G.~I. (2011).
\newblock {\em Encyclopedia of machine learning}.
\newblock Springer Science \& Business Media.

\bibitem[\protect\astroncite{Santoro et~al.}{2023}]{santoro2023analyzing}
Santoro, A., Galeazzi, A., Scantamburlo, T., Baronchelli, A., Quattrociocchi,
  W., and Zollo, F. (2023).
\newblock Analyzing the changing landscape of the covid-19 vaccine debate on
  twitter.
\newblock {\em Social Network Analysis and Mining}, 13(1):115.

\bibitem[\protect\astroncite{Schmidt et~al.}{2018}]{schmidt2018polarization}
Schmidt, A.~L., Zollo, F., Scala, A., Betsch, C., and Quattrociocchi, W.
  (2018).
\newblock Polarization of the vaccination debate on facebook.
\newblock {\em Vaccine}, 36(25):3606--3612.

\bibitem[\protect\astroncite{Stella et~al.}{2018}]{stella2018bots}
Stella, M., Ferrara, E., and De~Domenico, M. (2018).
\newblock Bots increase exposure to negative and inflammatory content in online
  social systems.
\newblock {\em Proceedings of the National Academy of Sciences},
  115(49):12435--12440.

\bibitem[\protect\astroncite{Zannettou et~al.}{2018}]{zannettou2018gab}
Zannettou, S., Bradlyn, B., De~Cristofaro, E., Kwak, H., Sirivianos, M.,
  Stringini, G., and Blackburn, J. (2018).
\newblock What is gab: A bastion of free speech or an alt-right echo chamber.
\newblock In {\em Companion Proceedings of the The Web Conference 2018}, pages
  1007--1014.

\bibitem[\protect\astroncite{Zannettou
  et~al.}{2019}]{zannettou2019disinformation}
Zannettou, S., Caulfield, T., De~Cristofaro, E., Sirivianos, M., Stringhini,
  G., and Blackburn, J. (2019).
\newblock Disinformation warfare: Understanding state-sponsored trolls on
  twitter and their influence on the web.
\newblock In {\em Companion proceedings of the 2019 world wide web conference},
  pages 218--226.

\end{thebibliography}
\clearpage
\section*{Supplementary Information}
\setcounter{figure}{0}
\setcounter{table}{0}
\begin{figure}[h]
    \centering
    \includegraphics[width=\textwidth]{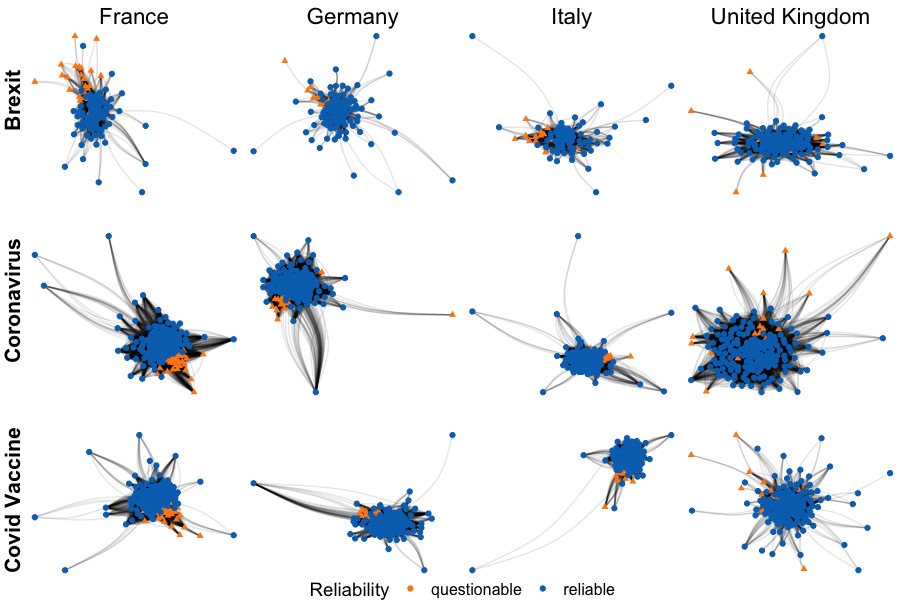}
    \caption{Similarity network among news outlets, where each news source is represented as a node, and edges represent audiences' similarity among news outlets. The color and shape of the nodes indicate the classification of the news source, and the thickness of the edges represents the level of similarity of retweeters between two news sources.  
    Each network represents the news outlets' similarity on one topic for one country.}
    \label{fig:complete_network}
\end{figure}

\begin{figure}[ht!]
    \centering
    \includegraphics[width=\textwidth]{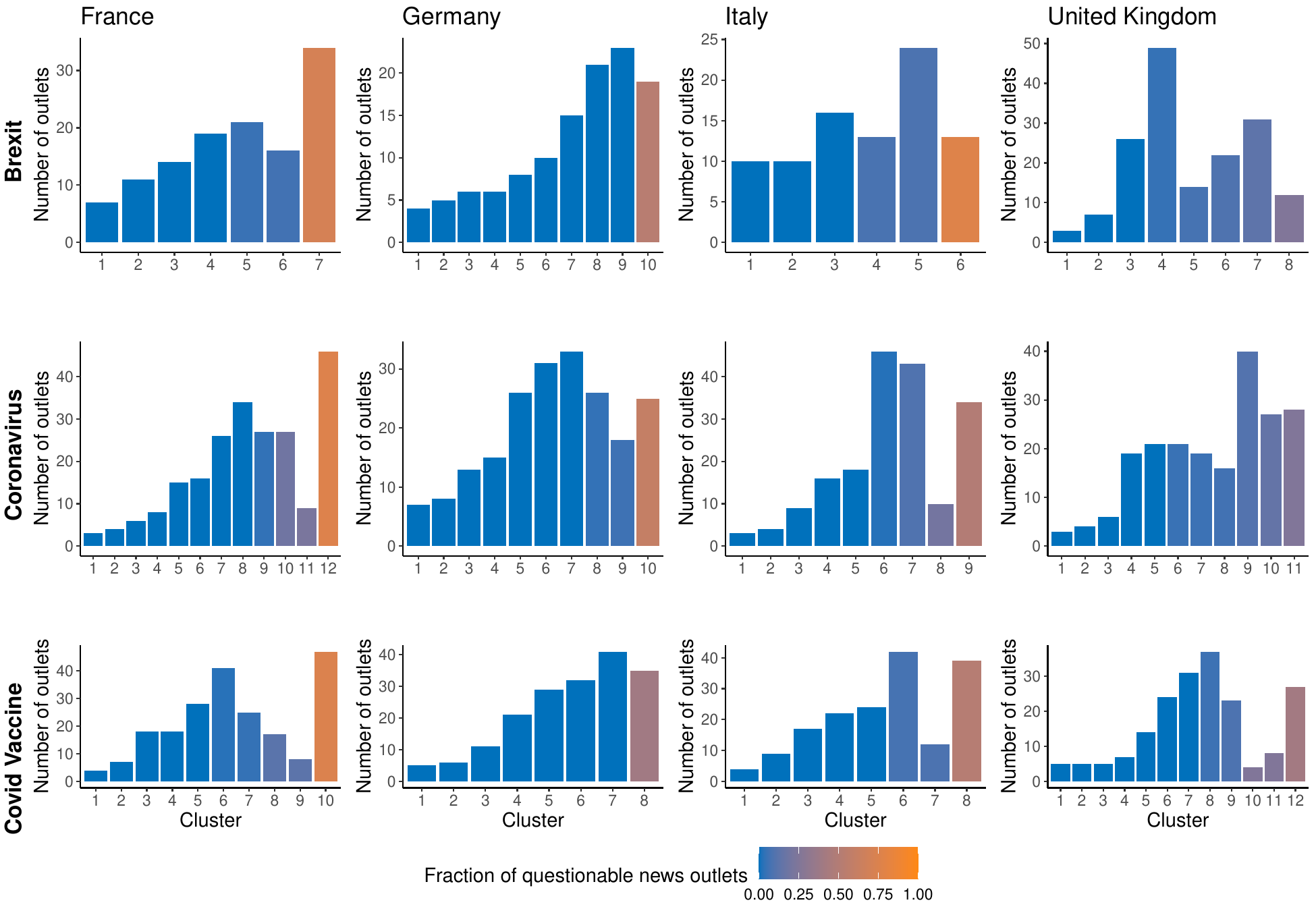}
    \caption{Community detection analysis of news outlets’ similarity networks with all the edges. Clusters were found using the Louvain clustering algorithm and sorted based on the percentage of questionable news outlets. The percentage of questionable sources in each cluster is color coded.}
    \label{fig:clusters_edges}
\end{figure}


\begin{figure}[ht!]
    \centering
    \includegraphics[width=\textwidth]{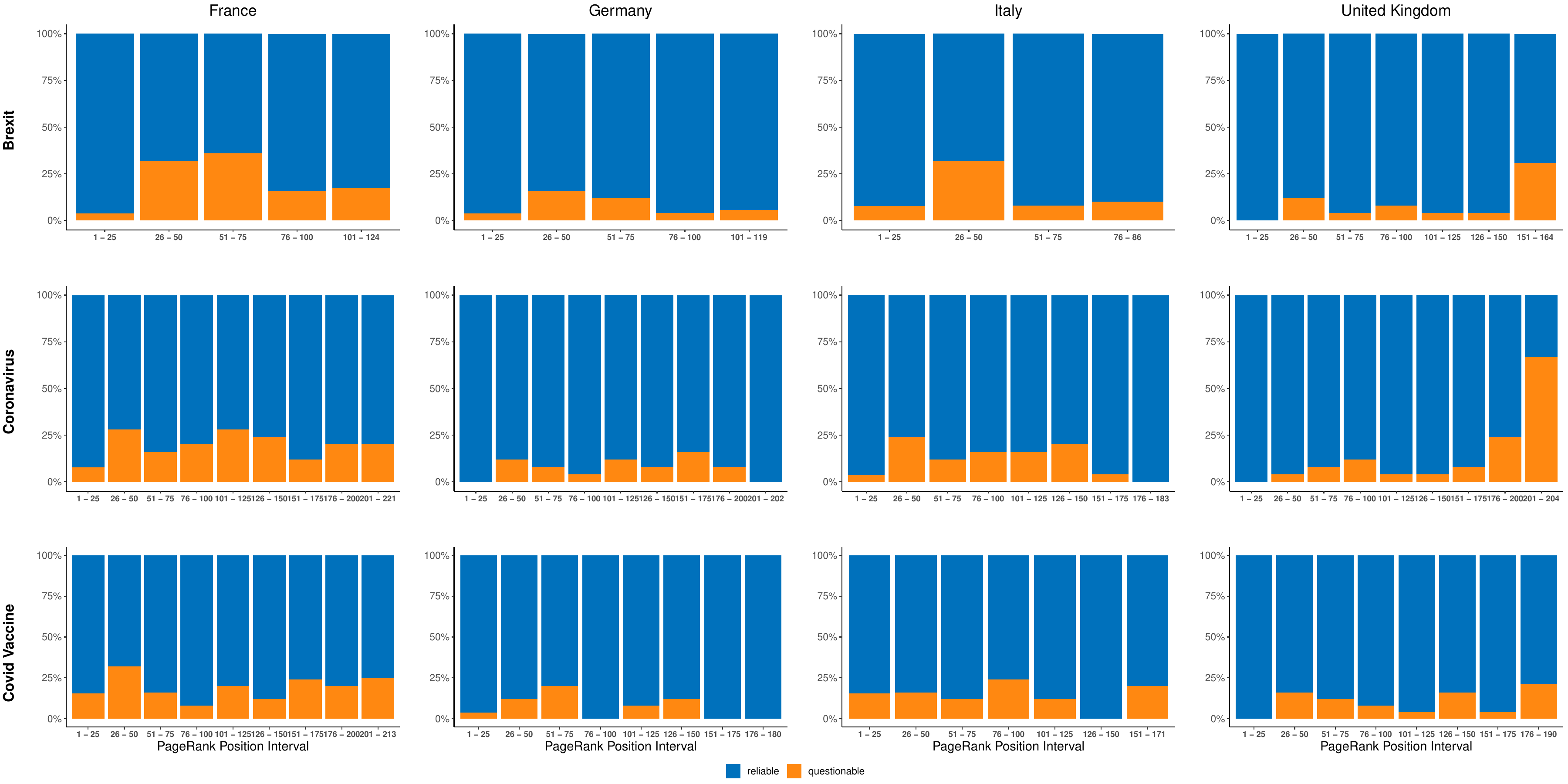}
    \caption{Distribution of News Outlets type respect to PageRank score}
    \label{fig:keywords_trends}
\end{figure}

\begin{table}[h]
    \centering
    \begin{tabular}{c|c|c|c|c}

    \hline
    \hline
    \textbf{Topic} & \textbf{France}  & \textbf{Germany} & \textbf{Italy} & \textbf{UK} \\ \hline

    Brexit & 0.41 & 0.64 & 0.48 & 0.05 \\ \hline

    Coronavirus & 0.25 & 0.48 & 0.25 & 0.18 \\ \hline

    Covid Vaccine & 0.24 & 0.44 & 0.21 & 0.22 \\

    \hline
    \hline
         
    \end{tabular}
    \caption{The table showcases adjusted assortativity coefficients \citep{karimi2022inadequacy} for key topics across France, Germany, Italy, and the UK. These coefficients measure the tendency of nodes to be connected to nodes with similar degrees within each country's topic-based network. Notably, variations across countries highlight distinct patterns of intra-network connectivity for each topic.}

    \label{tab:assortativity}
\end{table}






         


\begin{table}[h]
    \centering
    \begin{tabular}{c|c|c|c|c}

    \hline
    \hline
    \textbf{Topic} & \textbf{France}  & \textbf{Germany} & \textbf{Italy} & \textbf{UK} \\ \hline

    Brexit & 0.22 & 0.15 & 0.22 & 0.23 \\ \hline

    Coronavirus & 0.40 & 0.39 & 0.37 & 0.31 \\ \hline

    Covid Vaccine & 0.38 & 0.31 & 0.37 & 0.18 \\

    \hline
    \hline
         
    \end{tabular}
    \caption{Edge Density topic and country wise}
    \label{tab:edge_density}
\end{table}







         

\begin{table}[h]
    \centering
    \begin{tabular}{c|c|c|c|c}

    \hline
    \hline

    \textbf{Topic} & \textbf{Country} & \textbf{Connections} & \textbf{Total Edges} & \textbf{Percentage (\%)} \\ \hline

    \multirow{4}{*}{\textbf{Brexit}} & France & 268 & 1,684 & 15.92 \\ \cline{2-5}
    & Germany & 62 & 1,033 & 5.99 \\ \cline{2-5}
    & Italy & 112 & 801 & 13.98 \\ \cline{2-5}
    & UK & 266 & 3,088 & 8.62 \\ \hline

    \multirow{4}{*}{\textbf{Coronavirus}} & France & 2,389 & 9,764 & 24.47 \\ \cline{2-5}
    & Germany & 567 & 8,060 & 7.04 \\ \cline{2-5}
    & Italy & 1,162 & 6,230 & 18.63 \\ \cline{2-5}
    & UK & 610 & 6,465 & 9.43 \\ \hline

    \multirow{4}{*}{\textbf{Covid Vaccine}} & France & 2,066 & 8,546 & 24.13 \\ \cline{2-5}
    & Germany & 447 & 4,944 & 9.02 \\ \cline{2-5}
    & Italy & 1,143 & 5,422 & 21.08 \\ \cline{2-5}
    & UK & 410 & 3,263 & 12.54 \\

    \hline
    \hline
         
    \end{tabular}
    \caption{Connections between reliable and questionable news sources}
    \label{tab:connections}
\end{table}

\begin{table}[]
    \centering
    \begin{tabular}{c|c|c|c}
         \hline
         \hline

         \textbf{Topic} & \textbf{Country} & \textbf{Questionable Audience} & \textbf{Reliable Audience} \\ \hline

         \multirow{4}{*}{\textbf{Brexit}} & France & 3,808 & 31,310 \\ \cline{2-4}
         & Germany & 693 & 22,122 \\ \cline{2-4}
         & Italy & 1,860 & 7,407 \\ \cline{2-4}
         & UK & 18,074 & 225,227 \\ \hline

         \multirow{4}{*}{\textbf{Coronavirus}} & France & 56,404 & 449,682 \\ \cline{2-4}
         & Germany & 14,221 & 419,530 \\ \cline{2-4}
         & Italy & 18,624 & 143,624 \\ \cline{2-4}
         & UK & 38,204 & 900,928  \\ \hline

         \multirow{4}{*}{\textbf{Covid Vaccine}} & France & 75,953 & 368,751  \\ \cline{2-4}
         & Germany & 15,772 & 157,704  \\ \cline{2-4}
         & Italy & 27,348 & 146,502  \\ \cline{2-4}
         & UK & 23,951 & 293,756 \\

         \hline
         \hline
    \end{tabular}
    \caption{Audience count for reliable and questionable news sources}
    \label{tab:audience}
\end{table}
\end{document}